\documentclass[floatfix,aps,twocolumn,prl,superscriptaddress]{revtex4}
\usepackage{amsmath}
\usepackage{amssymb}
\usepackage{graphicx}
\usepackage{dcolumn}
\usepackage{natbib}
\usepackage{bm}
\usepackage{epstopdf}
\begin{document}
\title{Breakdown of the universal Josephson relation in spin ordered cuprate superconductors}
\author{A. A. Schafgans}
\email{aschafgans@physics.ucsd.edu}
\affiliation{Department of Physics, University of California, San Diego, La Jolla, California 92093, USA}
\author{C. C. Homes}
\affiliation{Condensed Matter Physics and Materials Science Department, Brookhaven National Laboratory, Upton, New York 11973, USA}
\author{G. D. Gu}
\affiliation{Condensed Matter Physics and Materials Science Department, Brookhaven National Laboratory, Upton, New York 11973, USA}
\author{Seiki Komiya}
\affiliation{Central Research Institute of the Electric Power Industry, Yokosuka, Kanagawa 240-0196, Japan}
\author{Yoichi Ando}
\affiliation{Institute of Scientific and Industrial Research, Osaka University , Ibaraki, Osaka 567-0047, Japan}
\author{D. N. Basov}
\affiliation{Department of Physics, University of California, San Diego, La Jolla, California 92093, USA}
\date{\today}

\begin{abstract}
We present \emph{c} axis infrared optical data on a number of Ba, Sr and Nd-doped cuprates of the La$_{2}$CuO$_{4}$ (La214) series in which we observe significant deviations from the universal Josephson relation linking the normal state transport (DC conductivity $\sigma_{DC}$ measured at $T_{c}$) with the superfluid density ($\rho_{s}$): $\rho_{s}\propto\sigma_{DC}(T_{c})$. We find the violation of Josephson scaling is associated with striking enhancement of the anisotropy in the superfluid density. The data allows us to link the breakdown of Josephson interlayer physics with the development of magnetic order in the CuO$_2$ planes.
\end{abstract}

\maketitle

Two decades of research in high transition temperature ($T_{c}$) superconductivity have uncovered universal behaviors that hold true for all cuprates \cite{Uemura-PRL62-2317-1989,Yamada-PRB57-6165-1998,Homes-Nature430-539-2004,Basov-PRB50-3511-1994,Dordevic-PRB65-134551-2002,Yu-NatPhys5-873-2009}. One example is a scaling relationship linking the normal state transport measured at $T_{c}$ ($\sigma_{DC}$($T_{c}$)) with the superfluid density ($\rho_{s}$): $\rho_{s}\propto\sigma_{DC}(T_{c})$ \cite{Dordevic-PRB65-134551-2002}. This relationship is general and holds both for the in-plane and interplane response. The universal relation, defined for the \emph{c} axis superfluid density and DC conductivity ($\rho^c_{s}\propto\sigma^c_{DC}(T_{c})$) originates from the profound connection between the collective response of the condensate below $T_c$ and single particle properties above $T_c$ in a layered superconductor. Ultimately, the superconducting condensate is formed at the expense of the normal state conductivity at $T_{c}$. The amount of spectral weight available for the condensate formation is therefore predetermined by the magnitude of $\sigma^c_{DC}(T_{c})$. This conjecture is particularly straightforward in those cases where the conductivity is weakly frequency dependent over the energy scale comparable to the superconducting gap (Fig.1 inset): a situation relevant to cuprates in the underdoped regime. Despite the fact that $\sigma^c_{DC}(T_{c})$ varies by 3-4 orders of magnitude between different families of cuprates, it allows for a remarkably reliable prediction of the magnitude of the c-axis superfluid density. In spite of this, small deviations from the $\rho^c_{s} \propto \sigma^c_{DC}(T_{c})$ trend are apparent on the overdoped side of the cuprate phase diagram \cite{Dordevic-PRB65-134551-2002}. These relatively minor departures can be accounted for by explicitly considering the energy scale involved in the condensate formation: $\rho^c_{s} \propto \sigma^c_{DC}(T_{c}) * T_{c}$ \cite{Homes-Nature430-539-2004}. With the addition of $T_{c}$, the universality of Josephson physics in cuprate superconductors has proven to be quite robust and therefore serious departures from this trend are significant. It is from this perspective that we present data demonstrating the unprecedented breakdown of this universal Josephson relationship in La214 materials.
\begin{figure}
\includegraphics[width=3.5in]{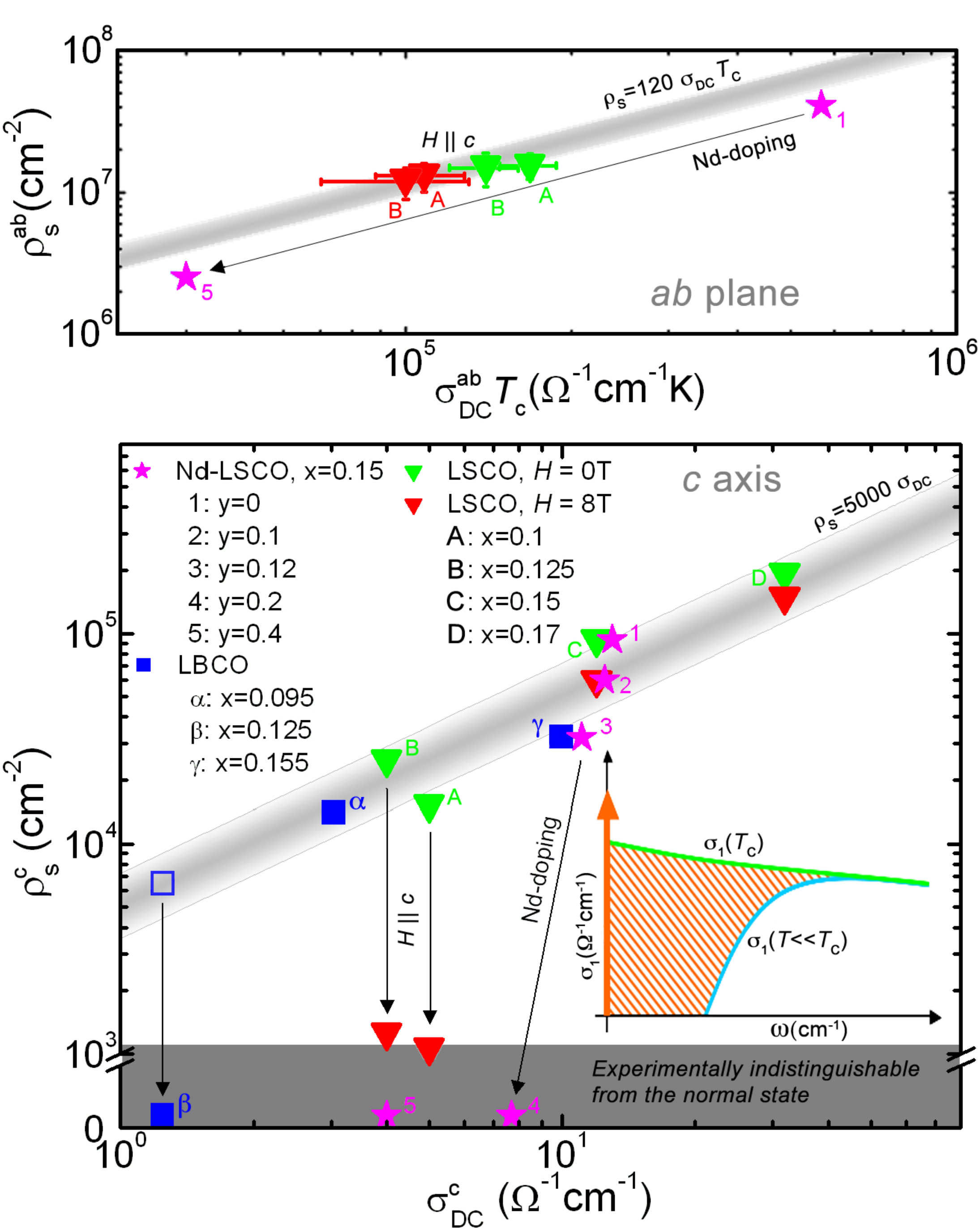}
\caption{\emph{Bottom}: The universal Josephson relation for the interlayer response of cuprate high $T_{c}$ superconductors $\rho_{s}^{c} \propto \sigma_{DC}^{c}(T_{c})$ (grey line). This relation breaks down as the result of applied field (LSCO, points A-D) and doping (Nd-LSCO, points 1-5 and LBCO, points $\alpha - \gamma$). The grey box at the bottom displays the lower limit of detectable superfluid density utilizing infrared optical techniques. Inset: Schematic of the normal state conductivity at $T_{c}$ (green curve) and the superconducting state conductivity (blue curve). The orange hashed region represents the spectral weight that is transferred from finite frequencies into a dissipationless superconducting delta peak (orange arrow) at zero frequency. In materials with weakly frequency dependent normal state conductivities, the DC conductivity provides an accurate estimate of the superfluid density. \emph{Top}: The in-plane universal relation holds in the samples where the \emph{c} axis Josephson relation breaks down.}
\label{Fig. 1}
\end{figure}
\begin{figure*}
\centering
\includegraphics[width=7in]{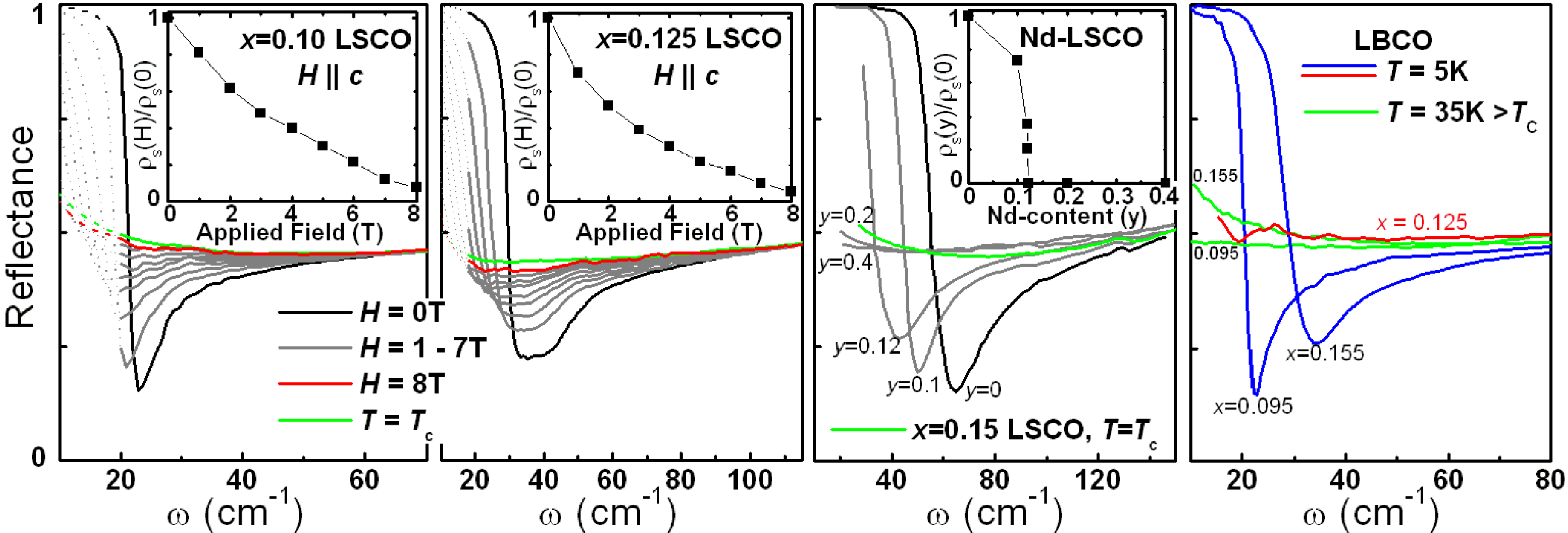}
\caption{Evolution of the Josephson plasma resonance with applied field and doping in a series of La-based single crystals. All data were taken at \emph{T}=8 K, except where specified. The Nd-LSCO data are replotted from \cite{Tajima-PRL86-500-2001}, however we measured the \emph{x}=0.15 LSCO $T_{c}$ curve on a separate crystal, shown here for reference. LBCO spectra shown for three dopings, demonstrating the absence of \emph{c} axis coherence in the \emph{x} = 0.125 sample. Insets: Normalized \emph{c} axis superfluid density as a function of applied field and Nd-content.}
\label{Fig. 2}
\end{figure*}

La$_{2-x}$Ba$_{x}$CuO$_{4}$ (LBCO) and La$_{2-x}$Sr$_{x}$CuO$_{4}$ (LSCO) samples in this study were grown using a traveling-solvent floating zone technique. The samples were taken from rods and the crystallographic axes were determined by Laue diffraction. The crystals were aligned to better than 1 degree on the Laue camera, then cut and polished with kerosene and diamond paste, with progressively finer and finer laps - the final lap is typically done with 0.1 micron diamond paste, yielding an optically flat surface with a bright finish.

The magnitude of the interlayer superfluid density was determined using IR spectroscopy. IR spectra measured with the polarization of the electric field vector normal to the CuO$_2$ planes (\emph{E} $\parallel$ \emph{c} axis) (Fig.2) allow one to register a collective plasma mode originating from Josephson coupling of the planes, (the Josephson plasma resonance (JPR)) \cite{Basov-PRB50-3511-1994,Matsuda-PhysC362-64-2001}. The superfluid density of the LSCO samples (Insets, Fig. 2 a, b) was determined by using an extrapolation-independent technique relying on the imaginary part of the complex optical conductivity ($\sigma_{2}$), as determined through Kramers-Kronig analysis. In the superconducting state, $\sigma_{2}$ is composed of a regular background component and a superconducting component $\sigma_{2}$ = $\sigma_{2}^R$ + $\sigma_{2}^S$. By taking into account the background contribution $\sigma_{2}^R$ as described in Ref. \cite{Dordevic-PRB65-134551-2002}, it is possible to accurately determine the superfluid density: $\rho_{s}$ = 4 $\pi \sigma_{2}^{S} \omega$. The model-independent analysis of the optical constants yields accurate values of $\rho_{s}^{c}$. 

Because the \emph{c} axis superfluid density is mediated by Josephson coupling, the JPR is sensitive to the phase relationship of the superconducting order parameter between neighboring planes:
\begin{equation}
\rho_{s}(F,T)=\rho_{s}(0,T)\langle cos(\phi_{n, n+1})\rangle .
\end{equation}
Here, $\langle cos(\phi_{n, n+1})\rangle$ represents the thermal and disorder average of the phase difference between layer n and n+1, and \emph{F} is a process that alters the superconducting phase relationship. Any process which, on average, produces a phase difference between layers and suppresses the superfluid density should be evident in the JPR.

The breakdown of the universal relationship occurs upon the application of an external magnetic field or doping, exemplified by a series of LSCO, LBCO, and La$_{1.85-y}$Nd$_{y}$Sr$_{0.15}$CuO$_{4}$ (Nd-LSCO) samples (Fig. 1, bottom). For low Nd content (stars 1-3), the interlayer superfluid density is reduced disproportionately faster compared to the universal trend. The complete breakdown of the universal relation is further evident once \emph{y} $>$ 0.12 (stars 4 and 5), in which any detectable sign of the \emph{c} axis superfluid density has vanished (Fig. 2 insets). Underdoped (UD) LSCO (triangles A and B, Fig. 1) reveals similar behavior in a $H \parallel c$ axis applied magnetic field. The behavior of the JPR is very sensitive to field orientation and for $H \parallel ab$ plane, even field values up to 17T are insufficient to significantly impact the JPR \cite{Dordevic-PRB71-054503-2005,Bentum-PhysicaC293-136-1997,Gerrits-PRB51-12049R-1995}. The \emph{c} axis superfluid density drastically decreases in a $H \parallel c$ axis field such that by \emph{H} = 8T, the \emph{c} axis response is indistinguishable from the normal state just above $T_{c}$. The DC conductivity is not impacted by the same field, which results in the 8T data points falling away from the universal line. On the contrary, the optimally doped (OP) LSCO samples (triangles C and D, Fig. 1) remain well described by the universal relation. The small suppression of the superfluid density in the OP-LSCO samples is fully accounted for by considering the role of vortices \cite{Schafgans-PRL104-157002-2010}. Importantly, for all fields and dopings discussed, the in-plane superfluid density remains well described by the universal relation (Fig. 1, top) \cite{Schafgans-PRL104-157002-2010,Tajima-EPL47-715-1999, Tajima-PRB71-094508-2005, Fujita-PRL95-097006-2005}, indicating that the superfluid density anisotropy $\gamma = \rho_{s}^{ab}/\rho_{s}^{c}$ becomes divergent. Later, we discuss the case of LBCO in relation to Fig. 1.

To understand the physics responsible for the breakdown of the Josephson relationship, Figure 2 shows the JPR in a number of La214 compounds as a function of applied field (LSCO, Figs. 2a, b), Nd-doping (Fig. 2c), and Ba-doping (Fig. 2d). In UD-LSCO, we observe the JPR to be quenched in moderate magnetic fields oriented $H \parallel c$ axis. We define the magnetic field required to quench the JPR below our experimental limitations and restore the \emph{c} axis IR spectra to the normal state values as the decoupling field $H_{D}$. The decoupling field corresponds to the vanishing of the superfluid density $\rho_{s}^{c}$ extracted from the optical conductivity \cite{Schafgans-PRL104-157002-2010}. Stripe-inducing co-doping \cite{Ichikawa-PRL85-1738-2000,Wakimoto-PRB67-184419-2003} can similarly quench the JPR. Results on a series of Nd-LSCO samples \cite{Tajima-PRL86-500-2001} and reproduced here (Fig. 2c), show that as Nd is added, the JPR is suppressed and moves to lower energies. Near a critical Nd-doping of \emph{y}=0.12, the JPR is completely quenched and the \emph{c} axis IR spectra return to the normal state values, indicating the \emph{c} axis superfluid has vanished. We note that the behavior of the JPR in UD-LSCO in a \emph{c} axis magnetic field is remarkably similar to what is observed in Nd-LSCO; however magnetic field provides the ability to continuously tune \emph{c} axis Josephson coupling within the same sample. By observing the loss of the JPR in these materials, we conclude that both a moderate \emph{c} axis magnetic field and stripe-inducing co-doping causes a drastic reduction of interlayer superconducting phase coherence.

Effects associated with the breakdown of Josephson coupling are further detailed in the phase diagrams shown in Figure 3. We plot the behavior of the decoupling field in \emph{x}=0.10 LSCO as a function of temperature in Fig. 3a. We observe the decoupling field to be below the in-plane resistive critical field $H_{c2}$ (black line) and our own in-plane optical measurements (not shown) indicate that the in-plane superfluid remains largely unaffected by the loss of Josephson coupling. The \emph{c} axis magnetic field causes a drastic reduction in interlayer superconducting coherence as seen in the loss of the superfluid density (Fig. 2 insets), at rates much faster than can be accounted for with standard vortex models (\cite{Schafgans-PRL104-157002-2010} and references therein). Therefore, below the decoupling field, the sample is in a three dimensional (3D) superconducting state characterized by Josephson coupled CuO$_{2}$ planes while above this field, the sample has transitioned into a two dimensional (2D) superconducting state. We obtained similar results for \emph{x}=0.125 LSCO, whereas the optimal (\emph{x}=0.15) and overdoped (\emph{x}=0.17) samples did not exhibit this behavior, with Josephson coupling remaining for all fields measured.
\begin{figure*}
\centering
\includegraphics[width=7in]{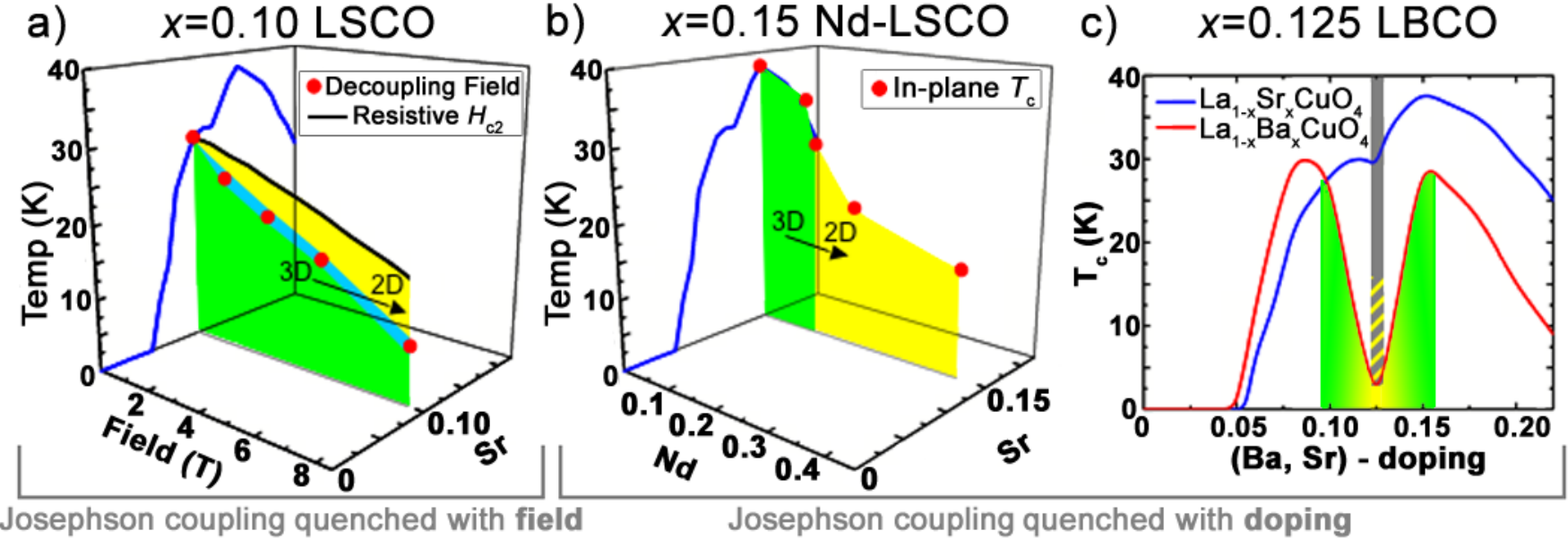}
\caption{Phase diagrams showing magnetic field and doping induced loss of interlayer Josephson coupling, in a number of La214 based materials. Here, the green regions labeled 3D represent parameter space of bulk superconductivity with prominent interlayer Josephson coupling. The yellow 2D regions represent the loss of interlayer Josephson coupling while in-plane superconductivity remains. In Fig. 3a, the black line is a constant contour of magneto-resistance near the $T_{c}$ value \cite{Lake-Nature415-299-2002} and the blue region represents error of the measured decoupling field. In Fig. 3c, the spin-stripe ordering transition temperature $T_{spin}$=40K (grey region) proceeds a Kosterlitz-Thouless transition to two-dimentional superconductivity \emph{T$_{KT}$}=16K (hashed bar), where the anisotropy of \emph{c} axis to \emph{ab} plane resistivity becomes infinite, within experimental uncertainty. At $T_{3D} \approx$ 4 K (yellow region) there is the existence of a bulk Meissner state, while Josephson coupling has not been observed.}
\label{Fig. 3}
\end{figure*}

Turning to Nd-LSCO, in Fig. 3b we replot $T_{c}$ as a function of Nd-doping \cite{Tajima-PRL86-500-2001}. Interestingly, $T_{c}$ is observed to experience a much smaller suppression as a function of Nd-doping than the \emph{c} axis superfluid density (Fig. 2 insets). This demonstrates that the in-plane superfluid remains intact inspite of the loss of Josephson coupling. Therefore, Nd-doping initiates a transition from the 3D superconducting state characterized by interlayer Josephson coupling to a 2D superconducting state where Josephson coupling is no longer observed. Even though the 3D-2D transition in UD-LSCO is more gradual than in Nd-LSCO, the net result is the same: the complete loss of interlayer Josephson coupling, in stark conflict with the expectations of the conventional theory of Josephson coupling.

Several attributes of the breakdown of the Josephson relationship are also apparent in La$_{1.875}$Ba$_{0.125}$CuO$_{4}$ (LBCO.125). Recently, Tranquada and Li \cite{Li-PRL99-067001-2007,Tranquada-PRB78-174529-2008} showed striking experimental results for LBCO.125 in which bulk $T_{c}$ is greatly suppressed from a similar doping $T_{c}(x=0.095)$ = 32 K to just $T_{3D}(x=0.125)\approx$ 4 K \cite{Moodenbaugh-PRB38-4596-1988}. Here, we show IR results demonstrating the behavior of the JPR through this doping range with suppressed $T_{c}$ (Fig. 2d). LBCO samples at nearby dopings show strong JPR features and are in full agreement with the universal Josephson relationship (Fig 1, blue squares $\alpha, \gamma$). Provided LBCO.125 complies with this universal relation, the expected value for the interlayer superfluid density can be inferred based on the \emph{c} axis conductivity (open blue square, Fig. 1) \cite{Li-PRL99-067001-2007,Tranquada-PRB78-174529-2008}. However, no evidence of the strong JPR feature corresponding to $\rho_{s}^{c} = 7000 cm^{-2}$ is found in the IR data for this compound (Fig. 2d). Indeed, LBCO.125 at low temperatures is identical to the $T \gtrsim T_{c}$ spectra at nearby dopings.

These observations point to anomalously anisotropic superconductivity in LBCO.125, similar to Nd-LSCO and UD-LSCO in magnetic field. We have represented LBCO.125 data in Fig. 3c. The vertical bar at \emph{x}=1/8 schematically shows the transport results where below the spin-stripe ordering transition temperature $T_{spin}$=40 K (grey region), LBCO.125 exhibits behavior reminiscent of the onset of in-plane superconductivity. At zero field, Li \emph{et. al.} identified $T_{spin}$=40K as the most likely onset temperature of in-plane superconductivity \cite{Li-PRL99-067001-2007}, supported by angle-resoved photoemission data demonstrating the presence of a gap consistent with \emph{d}-wave symmetry \cite{Valla-Science314-1914-2006}. For Fig. 1, $T_{spin}$ is the temperature we used to determine $\sigma_{DC}$. Importantly, \emph{c} axis resistivity becomes immeasurably small near \emph{T}=10 K, yet below this temperature we do not observe Josephson coupling \cite{Homes-PRL96-257002-2006}.

We note again that the in-plane behavior of the superfluid remains well described by the universal relation (Fig. 1, top) even when \emph{c} axis Josephson coupling vanishes. Based on the sensitivity of our experimental setup and difficulty of infrared in-plane superfluid measurements in cuprates, we can bound the amount of remaining superfluid density: in UD-LSCO at \emph{T}=8K, at least 70$\%$ of the in-plane superfluid remains at \emph{H}=8T. Data available in the literature on Nd-LSCO \cite{Tajima-EPL47-715-1999} demonstrates that the in-plane superfluid can be determined by $\sigma_{DC}(T_{c})*T_{c}$, regardless of the loss of c-axis coherence.

What we observe in the cases presented here is the complete breakdown of Josephson interlayer coupling, induced by an applied magnetic field, Nd or Ba-doping. Ostensibly, these processes are very different. However, they do share one common attribute: all these processes are known to stabilize and enhance static, long-range magnetic order \cite{Tajima-EPL47-715-1999,Tranquada-Nature375-561-1995,Khaykovich-PRB71-220508-2005,Valla-Science314-1914-2006,Lake-Nature415-299-2002,Sonier-PRB76-064522-2007,Chang-arXiv-07122181v2,Machtoub-PRL94-107009-2005,Savici-PRL95-157001-2005}. Significantly, all three materials discussed here exhibit in-plane magnetic order that is otherwise not present at nearby doping levels or in the absence of an applied field. Here we underscore the notion that in-plane magnetic order preferentially alters collective pair tunneling along the \emph{c} axis while the in-plane superfluid remains relatively unimpacted.

The bilayer cuprate YBa$_{2}$Cu$_{3}$O$_{y}$ (YBCO), subjected to an external magnetic field applied in the same manner as the LSCO samples presented here, does not display as drastic a loss of Josephson coupling as seen in LSCO \cite{LaForge-PRL101-097008-2008,LaForge-PRB76-054524-2007}. There is nonetheless an appreciable reduction in the \emph{c} axis superfluid density for an applied field for UD YBCO samples, however the rate of the reduction is in line with the vortex wandering model and does not exhibit any anomalous behavior from the standpoint of the universal Josephson relation. The measurements were performed on samples doped between y = 6.67 and 6.95, corresponding to hole concentrations between x =  0.12 and 0.18 \cite{Liang-PRB73-180505R-2006}. Yet even for samples doped near the 1/8 hole concentration where spin order is thought to exist \cite{Daou-Nature463-519-2010}, static magnetic order is not observed in YBCO until temperatures much below our measurements \cite{Hinkov-Science319-597-2008}, implying that YBCO lacks the appropriate magnetic order to facilitate the suppression of \emph{c} axis Josephson coupling.

Attempts to theoretically understand the breakdown of Josephson coupling have been presented by two groups \cite{Berg-PRL99-127003-2007,Berg-arXiv09014826,Berg-NatPhys5-830-2009,Wollny-arXiv09075202}. Both scenarios posit that charge and spin-density wave (CDW and SDW) stripe order suppresses interlayer Josephson coupling. Based on the experimental discoveries discussed here, we can identify several additional aspects of a complete theory of dynamical layer decoupling. In addition to LBCO.125 and Nd-LSCO, the observation of magnetic-field induced phase decoherence in UD-LSCO creates a significantly more stringent set of experimental constraints on such a theory. Namely, a complete theory cannot rely on commensurability and must be applicable to both the low temperature tetragonal and the low temperature orthorhombic structures of LSCO and LBCO. Since CDW order has not been experimentally observed in UD-LSCO, SDW stripe order appears to be more salient to interlayer decoherence. Additionally, effects only associated with doping and not moderate magnetic fields, such as a modified band structure, cannot be relied on as the sole mechanism of the Josephson breakdown. Finally, based on Fig. 1, we observe that the process of applying a magnetic field or doping only destroys pair tunneling and does not seem to impact the single particle properties of the normal state.

We thank S.A. Kivelson, E. Fradkin, J.M. Tranquada, and E. Berg for great discussions and acknowledge funding from the NSF and AFOSR MURI. Y. Ando was supported by KAKENHI 19674002 and 20030004.

\end{document}